\documentclass[preprint,a4paper]{article}

\usepackage{natbib}

\usepackage{amssymb}
\usepackage{amsfonts}
\usepackage{latexsym}
\usepackage[T1,T2A]{fontenc}
\usepackage[cp1251]{inputenc}
\usepackage[T2B]{fontenc}
\usepackage[bulgarian,english]{babel}

\newtheorem{theorem}{\bf Theorem}[section]
\newtheorem{algorithm}{\bf Algorithm}
\newtheorem{proposition}{\bf Proposition}[section]

\title{On some randomized algorithms and their evaluation}
\author{Krasimir Yordzhev}
\date{}

\begin{document}
\inputencoding{cp1251}

\maketitle
\begin{center} {\em
Trakia University\\ 
Stara Zagora, Yambol, Bulgaria} \\
E-mail: krasimir.yordzhev@gmail.com
\end{center}

\begin{abstract}

The paper considers  implementations of some randomized algorithms in connection with obtaining a random $n^2 \times n^2$ Sudoku matrix with programming language C++. For this purpose we describes the set $\Pi_n$ of all $(2n) \times n$ matrices, consisting of elements of the set $\mathbb{Z}_n =\{ 1,2,\ldots ,n\}$, such that every row is a permutation. We emphasize the relationship between these matrices and the $n^2 \times n^2$ Sudoku matrices. An algorithm to obtain random $\Pi_n$ matrices  is presented in this paper. Several auxiliary algorithms that are related to the underlying problem have been described.  We evaluated all algorithms according to two criteria - probability evaluation, and time for generation of random objects and checking of belonging to a specific set. This evaluations are interesting from both theoretical and practical point of view because they are particularly useful in the analysis of computer programs.

\end{abstract}

Keyword: {\it randomized algorithm, random object, permutation, binary matrix, algorithm evaluation, Sudoku matrix}

MSC[2010] code: 05B20, 65C05 68W40

\section{Introduction}\label{intr}

This article is intended for anyone who studies programming, as well as for his teachers. The work is a continuation and addition of  \citep{IJMSEA}. Here, along with the basic definitions and ideas for constructing randomized algorithms outlined in the  cited above publication, we will also describe specific implementations of these algorithms in the C++ programming language.

To demonstrate the ideas outlined in the article, we have chosen the C ++ programming language \citep{Magda, Magda_OOP, Magda_PSD}, but they can be implemented in any other algorithmic language. We hope that students who prefer to write in Java \citep{hadzhikolevi} or any other modern programming language will have no problems with the implementation of the algorithms we have proposed.

The presented in the article source codes have been repeatedly tested with various input data and they work correctly.

Let $\mathfrak{M}$ be a finite set. \emph{A Random objects generator of $\mathfrak{M}$}  is every algorithm $\mathcal{A}_\mathfrak M$  randomly generating any element of $\mathfrak M$, while elements generated by a random objects generator will be called \emph{random elements } of $\mathfrak M$, i.e. random numbers, random matrices, random permutations, etc. We take for granted that probabilities to obtain different random elements of  $\mathfrak M$ by means of $\mathcal{A}_\mathfrak M$ are equal, and are also equal to $\displaystyle \frac{1}{\left| \mathfrak{M} \right|}$. We denote the time that the random objects generator needs to obtain a random element of $\mathfrak M$ with $T(\mathcal{A}_\mathfrak{M} )$.

By \emph{randomized algorithm} we will mean any algorithm  which essentially uses a random object generator in its work.

The randomized algorithms are very often used to solve problems, which are proved to be NP-complete. For detailed information about NP-complete problems and their application see \citep{garey} or \citep{hopcroft}. A proof that a popular Sudoku puzzle is NP-complete is given in \citep{yato} and \citep{yatoseta}.

In this study, we will solve some particular cases from the following class of problems:

Let $n$ and $m=m(n)$ be natural numbers. Let us consider the set $\mathcal{U}$, consisting of objects every of which dependent on $m$ parameters, and every parameter belongs to the finite set $\mathfrak M$. We assume that there is a rule that uniquely describes object $u\in\mathcal{U}$, if all $m$ parameters are specified. Let $\mathcal{V} \subset \mathcal{U}$. The problem is to obtain (at least one) object, which belongs to the set $\mathcal{V}$. The number of the elements of the sets $\mathcal{U}$ and $\mathcal{V}$ depends only on the parameter $m$, which is an integer function of the argument $n$.

The standard algorithm that solves the above problem is briefly described as follows:

\begin{algorithm}\label{alg1}   \hfill

1) We obtain consequently $m=m(n)$ random elements of $\mathfrak M$ using random objects generator $\mathcal{A}_\mathfrak M$ and so we get the object $u\in\mathcal{U}$;

2) We check if $u\in\mathcal{V}$. If the answer is no, everything is repeated.
\end{algorithm}

In other words, if we already have a random objects generator, a randomized algorithm can be used as a generator of more complex random objects. 

The efficiency of Algorithm \ref{alg1} depends on the particular case in which it is used and can be evaluated according to the following criteria:

\textbf{Probability evaluation}: If $p(n)$ denotes the probability after generating $m=m(n)$ random elements of $\mathfrak{M}$ of obtaining an object of $\mathcal{V}$, then according to the classical probability formula:
\begin{equation}\label{p(n)}
p(n) = \frac{\left| \mathcal{V}\right|}{\left| \mathcal{U}\right|}
\end{equation}

\textbf{Time for generating and checking}: We denote by $\tau (n)$ the time needed to execute one iteration (repetition) of Algorithm \ref{alg1}. Then
\begin{equation}\label{tau}
\tau (n) =m(n) T(\mathcal{A}_\mathfrak{M})+\theta (n) ,
\end{equation}
where $\theta (n)$ is the time to examine if the obtained object belongs to the set $ \mathcal{V}$.

It is obvious that the efficiency of Algorithm \ref{alg1} will be proportional to $p(n)$ and inversely proportional to $\tau (n)$.

Of course, time for generating and checking does not give us the total time to execute the algorithm, since the number of repetitions is not known in advance. However, the characteristic $\tau (n)$ is essential to the effectiveness of any randomized algorithms.

The cases in which probability evaluation is equal to 1, i.e. the cases in which the algorithm is constructed directly to obtain element of the set $\mathcal V$ and there is no need of belonging examination, are of great interest, as only one iteration is implemented then, i.e. there is no repetition. 

Let $n$ be a positive integer. We denote by $\mathbb{Z}_n$ the set of the integers
$$\mathbb{Z}_n =\left\{ 1,2,\ldots ,n \right\} .$$

There are standard procedures for obtaining random numbers of the set $\mathbb{Z}_n$ in most of the programming environments. We take this statement for granted and we will use it in our examinations. Let $\mathcal{A}_n$ be a similar procedure. In the current study, we will consider that for $n\ne l$
\begin{equation}\label{eqt0}
T (\mathcal{A}_n )\approx T (\mathcal{A}_l) \approx t_0 = Const  .
\end{equation}

Below we show an example of a C ++ function that generates a random positive integer belonging to the set $\mathbb{Z}_n =\{ 1,2,\ldots ,n\}$:

\begin{algorithm} \hfill
\begin{verbatim}
int rand_Zn(int n)
{
    return rand() % n + 1;
}
\end{verbatim}
\end{algorithm}

In order for the function \verb"rand_Zn(int)"  to work so that every time we execute the program in which we will use it to obtain various random numbers, we must add the procedure
\begin{verbatim}
    srand(time(0));
\end{verbatim}
before first accessing this function, for example, at the beginning of the  \verb"main()" function. The functions \verb"rand()" and \verb"srand(s)" are from the library \verb"<cstdlib>" and the function \verb"time(t)" is from the library \verb"<ctime>". For more details, see for example \citep[p. 75]{Azalov_Zlatarova}.

Let $P_{ij}$, $0\le i,j\le n-1$ be $n^2$ in number square $n\times n$ matrices, whose elements belong to the set $\mathbb{Z}_{n^2} =\{ 1,2,\ldots ,n^2 \}$. Then $n^2 \times n^2$ matrix
$$
P = \left[ P_{ij} \right] =
\left[
\begin{array}{cccc}
P_{0\; 0} & P_{0\; 1} & \cdots & P_{0\; n-1} \\
P_{1\; 0} & P_{1\; 1} & \cdots & P_{1\; n-1} \\
\vdots & \vdots & \ddots & \vdots \\
P_{n-1\; 0} & P_{n-1\; 1} & \cdots & P_{n-1\; n-1}
\end{array}
\right]
$$
is called a \emph{Sudoku matrix}, if every row, every column and every submatrix $P_{ij}$, $0\le i,j\le n-1$ make \emph{permutation} of the elements of set $\mathbb{Z}_{n^2}$, i.e. every number $s\in \{ 1,2,\ldots ,n^2 \}$ is present only once in every row, every column and every submatrix $P_{ij}$. Submatrices $P_{ij}$ are called blocks of $P$.

In this paper we will illustrate the above mentioned ideas by analyzing some randomized algorithms for obtaining an arbitrary permutation of $n$ elements, an arbitrary $n^2 \times n^2$ Sudoku matrix  and an arbitrary $(2n) \times n$ matrix  with $2n$ rows and $n$ columns, every column of which is a permutation of $n$ elements.

We will prove that the problem for obtaining ordered $n^2$ - tuple of $(2n) \times n$ matrices, every row of which is a permutation of elements of $\mathbb{Z}_n$ is equivalent to the problem of generating a Sudoku matrix. We will analyze some possible algorithms for generating a random Sudoku matrix.

 How to create computer program for Sudoku solving (a mathematical model of the algorithm), using the concept set combined with the trial and error method is described in  \citep{AzBouki}.

\section{Random permutations}\label{sec2}
\par Let $n$ be an positive integer. We denote by $\mathcal{S}_n$ the set of all \emph{permutations} $\langle a_0 ,a_1 ,\ldots ,a_{n-1} \rangle$, where $a_i \in \mathbb{Z}_n$ and $a_i \ne a_j$ when $i\ne j$, $0\le i,j\le n-1$.

If $\sigma = \langle a_0 ,a_2 ,\ldots ,a_{n-1} \rangle \in \mathcal{S}_n$ is a permutation of all elements of the set $\mathbb{Z}_n =\{ 1,2,\ldots ,n\}$ then obviously $\sigma$ depends of $m(n)=n$ parameters $a_0$, $a_1$, $\ldots ,$ $a_{n-1}$.

As is well known, the number of all $n$-tuples of integers $\langle x_1 , x_2 ,\ldots ,x_n \rangle$, $x_i \in \mathbb{Z}_n$ is equal to
\begin{equation}\label{n-tuples}
|\underbrace{\mathbb{Z}_n \times \mathbb{Z}_n \times \cdots \times \mathbb{Z}_n}_n |=n^n
\end{equation}
and the number of all permutations of n elements is equal to
\begin{equation}\label{numbpermut}
|\mathcal{S}_n |=n!=1\cdot 2\cdot 3\cdots n .
\end{equation}

We denote by $p_1 (n)$ the probability to obtain a random permutation of $\mathcal{S}_n$ with the help of Algorithm \ref{alg1}. Then according to equations  (\ref{p(n)}), (\ref{n-tuples}) and (\ref{numbpermut}) we obtain:

\begin{equation}\label{p1}
p_1 (n) = \frac{n!}{n^n}
\end{equation}

The next algorithm works in time $O(n)$ and checks if ordered $n$-tuple $\langle a_0 ,a_1 ,\ldots a_{n-1} \rangle$, $a_i \in \mathbb{Z}_n$, $i=0,1,\ldots ,n-1$ is a permutation.

\begin{algorithm}\label{alg2}
Check if given integer array \verb"a[n]"  is a permutation of the elements from the set $\mathbb{Z}_n =\{ 1,2,\ldots ,n\}$.

\begin{verbatim}
bool alg3(int a[], int n)
{
    int v[n+1];
    for (int i=1; i<=n; i++) v[i] = 0;
    for (int i=0; i<n; i++)
    {
        if ( (a[i] > n) || (a[i] < 1) ) return false;  /* \end{verbatim} \rm Incorrect data - there is an element that does not belong to the set $\mathbb{Z}_n$ \begin{verbatim} */
        v[a[i]]++;
        if (v[a[i]] > 1)  return false;  /* \end{verbatim} \rm  Since the number \verb"a[i]" occurs more than once in the $n$-tuple. \begin{verbatim} */
    }
    return true;
}
\end{verbatim}

\end{algorithm}

Let $\tau_1 (n)$ denote the time for implementing one iteration of Algorithm \ref{alg1} when a random permutation of elements of $\mathbb{Z}_n$ is obtained, and let $\theta_1 (n) $ denote the time for checking whether an arbitrary $n$-tuples of numbers of $\mathbb{Z}_n $ belongs to $\mathcal{S}_n$. Analyzing Algorithm \ref{alg2}, it is easy to see that
\begin{equation}\label{pprop1}
\theta_1 (n) \in O(n) .
\end{equation}

Then, having in mind the equations (\ref{tau}), (\ref{eqt0}) and  (\ref{pprop1}) we obtain the following time for generating and checking:
\begin{equation}\label{tau1}
\tau_1 (n) =n T(\mathcal{A}_n)+\theta_1 (n) \in nt_0 +O(n)=O(n).
\end{equation}

The following algorithm is also randomized  (random integers are generated), but its probability evaluation is equal to 1, i.e. in Algorithm \ref{alg1} step 2 is not implemented, because when the first random $n$ numbers are generated the obtained ordered $n$-tuple is a permutation.

\begin{algorithm}\label{alg3} Obtaining random permutation $\sigma =\langle a_0 ,a_1 ,\ldots ,a_{n-1} \rangle \in \mathcal{S}_n ,$ where $a_i \in \mathbb{Z}_n$, $i=1,2,\ldots ,n$, $a_i \ne a_j$ when $i\ne j$.

\begin{verbatim}
void alg4(int a[], int n)
{
    int v[n];
    for (int i=0; i<n; i++) v[i] = i+1;
    int r;
    for (int i=0; i<n; i++)
    {
        r = rand_Zn(n-i)-1;
        a[i] = v[r];  /* \end{verbatim}
{\rm We remove the element $v[x]$ and reduce the number of the elements of the array with 1.} \begin{verbatim} */
        for (int j=r; j<=n-i; j++) v[j] = v[j+1];
    }
}
\end{verbatim}

\end{algorithm}

Analyzing the work of Algorithm \ref{alg3}, we see that in the end, $n$ different elements of the set $\mathbb{Z}_n =\{ 1,2,\ldots ,n\}$ are filled in the array \verb"a[n]". We consistently randomly take these elements from the array \verb"v" with length $n$ where all integers $1,2,\ldots , n$ are filled, such that \verb"v[i] = i+1", $i=0,1,\ldots ,n-1$. After each choice we remove the  selected integer from the array \verb"v", i.e. from the integers which we will randomly take in the next steps of the algorithm. Therefore Algorithm \ref{alg3} which obtains random permutation has probability evaluation:
\begin{equation}\label{p2}
p_2 (n) =1
\end{equation}

In Algorithm \ref{alg3}, there are two nested loops, so that  the outer loop is repeated exactly $n$ times. Internal loop in the worst case (if randomly selected integer equal to the value of the item  \verb"v[0]") will be repeated as many times as is the length of the array \verb"v", which in the first iteration of the outer loop is equal to $n$ and decreased each time by 1. Thus, we obtain the following time evaluation for  generating and checking in one iteration of Algorithm \ref{alg3}.
\begin{equation}
\tau_2 (n) \in t_0 \left[ O(n) +O(n-1)+\cdots +O(1)\right] =O(n^2 ) .
\end{equation}

We see that Algorithm \ref{alg1} applied for obtaining random permutations is more efficient than Algorithm \ref{alg3} in terms of the time for generating and checking. But on the other hand, the probability evaluation of Algorithm \ref{alg3} is equal to $p_2 (n) =1$.  The  probability evaluation of Algorithm \ref{alg1} is equal to $\displaystyle p_1 (n) = \frac{n!}{n^n} < 1$ for $n\ge 2$. Considering that the number of iterations is previously unknown and $\displaystyle \lim_{n\to\infty} p_1 (n) =0$, then  this makes Algorithm \ref{alg3} overall much more effective than the Algorithm \ref{alg1} when applied to generate random permutations.

\section{Random $(2n)\times n$ matrices, every row of which is a permutation of elements of $\mathbb{Z}_n$ }\label{sec3}

Let $\Pi_n$ denote the set of all $(2n)\times n$ matrices, which are also called $\Pi_n$ matrices, in which every row is a permutation of all elements of $\mathbb{Z}_n$. In this case $\mathfrak{M}=\mathbb{Z}_n$ and $m(n)=2n^2$. It is obvious that
\begin{equation}\label{Pi_n}
\left| \Pi_n \right| =\left( n! \right)^{2n}
\end{equation}

It is easy to see that when we obtain random $\Pi_n$ matrix with the help of Algorithm \ref{alg1} the following evaluations can be observed:

Probability evaluation:
\begin{equation}\label{p3(n)}
p_3 (n) = \frac{\left| \mathcal{V}\right|}{\left| \mathcal{U}\right|} =\frac{(n! )^{2n}}{n^{2n^2}}
\end{equation}

Time for generating and checking:
\begin{equation}\label{tau3}
\tau_3 (n) =m(n) T(\mathcal{A}_n)+2n\tau_1 (n)\in 2n^2 t_0 +2nO(n)=O(n^2 ) ,
\end{equation}
where $\tau_1 (n)$ is obtained according to equation (\ref{tau1}).

The next randomized algorithm will be more efficient than Algorithm \ref{alg1} in obtaining random $\Pi_n$ matrix according to the probability evaluation.

\begin{algorithm}\label{alg4}
Obtaining a random matrix $M =\left[ \pi_{ij}\right]_{2n\times n}\in \Pi_n$ with probability evaluation equal to 1. The algorithm will write the elements of the matrix $M =\left[ \pi_{ij}\right]_{2n\times n}$ in the array \verb"p[]" with size  $2n^2$, such that $\pi_{ij} = $\verb"p[i*n+j]", where $0\le i<2n$, $0\le j<n$.

\begin{verbatim}
void alg5(int p[], int n)
{
    int m=2*n;
    int a[n];
    for (int i=0; i<m; i++)
    {
        alg4(a,n);
        for (int j=0; j<n; j++)
        {
            p[i*n+j] = a[j];
        }
    }
}
\end{verbatim}

\end{algorithm}

Practically, Algorithm \ref{alg4} repeats Algorithm \ref{alg3} $2n$ times. As  Algorithm \ref{alg3} has a probability evaluation equal to 1, then Algorithm \ref{alg4} which obtains random $\Pi_n$ matrix also has probability evaluation
\begin{equation}\label{p4}
p_4 (n) =1
\end{equation}

It is easy to see that Algorithm \ref{alg4} works in time for generating and checking
\begin{equation}\label{tau4}
\tau_4 (n) = 2n\tau_3 (n) \in 2nO(n^2 ) =O(n^3 ) .
\end{equation}

As we can see below $\Pi_n$ matrices can successfully be used to create algorithms that are efficient in developing Sudoku matrices.

\section{S-permutation matrices}

A \emph{binary} (or \emph{boolean}, or \emph{(0,1)-matrix}) is called a matrix whose elements belong to the set ${\mathfrak B}=\{ 0,1\} $.  With ${\mathfrak B}_{n\times m}$ we will denote the set of all $n\times m$ binary matrices. The set of all square $n\times n$ binary matrices we will denote with ${\mathfrak B}_n ={\mathfrak B}_{n\times n}$

Two binary   matrices $A=\left[a_{ij} \right]_{n\times m} \in \mathfrak{B}_{n\times m}$ and $B=\left[ b_{ij} \right]_{n\times m} \in \mathfrak{B}_{n\times m}$ will be called \emph{disjoint} if there are not elements $a_{ij} \in A$ and $b_{ij} \in B$ with one and the same indices $i$ and $j$, such that $a_{ij} =b_{ij} =1$, i.e. if $a_{ij} =1$ then $b_{ij} =0$ and if $b_{ij} =1$ then $a_{ij} =0$, $1\le i\le n$, $1\le j\le m$.

A square binary matrix  $A\in {\mathfrak B}_n$ is called \emph{ permutation}, if there is only one 1 in every row and every column of the matrix $A$.

Let $\Sigma_{n^2}$ denote the set of all permutation $n^2 \times n^2$ matrices of the following type
\begin{equation}\label{matrA}
A =
\left[
\begin{array}{cccc}
A_{0\; 0} & A_{0\; 1} & \cdots & A_{0\; n-1} \\
A_{1\; 0} & A_{1\; 2} & \cdots & A_{1\; n-1} \\
\vdots & \vdots & \ddots & \vdots \\
A_{n-1\; 0} & A_{n-1\; 1} & \cdots & A_{n-1\; n-1}
\end{array}
\right] ,
\end{equation}
where for every $s,t\in \{ 0,1,2,\ldots ,n-1\}$  $A_{st}$ is a square $n\times n$ binary submatrix (\emph{block}) with only one element equal to 1, and the rest of the elements are equal to 0. The elements of $\Sigma_{n^2}$ will be called \emph{S-permutation}.

Geir Dahl introduces the concept of S-permutation matrix \citep{dahl} in relation to the popular \emph{Sudoku puzzle} giving the following obvious proposition:

\begin{proposition}\label{disj} \citep{dahl}
A square $n^2 \times n^2$ matrix $P$ with elements of  $\mathbb{Z}_{n^2} =\{ 1,2,\ldots ,n^2 \}$ is Sudoku matrix if and only if there are mutually disjoint matrices $A_1 ,A_2 ,\ldots ,A_{n^2} \in\Sigma_{n^2}$ such that $P$ can be presented as follows:
$$P=1\cdot A_1 +2\cdot A_2 +\cdots +n^2 \cdot A_{n^2}$$
\hfill $\Box$
\end{proposition}

As it is proved in \citep{dahl} and with other methods in \citep{YORDZHEV20133072}
\begin{equation}\label{sssssss}
\left| \Sigma_{n^2} \right| =\left( n! \right)^{2n}
\end{equation}

Therefore, if a random $\Sigma_{n^2}$ matrix obtained by means of Algorithm \ref{alg1} the following probability evaluation is valid:

\begin{equation}\label{ppp5}
p_5 (n)=\frac{(n!)^{2n}}{2^{(n^2 )^2}} =\frac{(n!)^{2n}}{2^{n^4}}
\end{equation}

In order to obtain a random $\Sigma_{n^2}$ matrix, we have to generate $m(n)=(n^2 )^2 =n^4$ random integers, which belong to the set $\{ 0,1\}$. Hence, whatever randomized algorithm  is used, the result is the following time for generating and checking:
\begin{equation}\label{ttt5}
\tau_5 (n) =m(n)T(\mathcal{A}_2) +\theta_5 (n)=n^4 t_0 +\theta_5 (n) \in O(n^z ),\quad z\ge 4,
\end{equation}
where $\theta_5 (n)$ is time for checking that the given binary matrix belongs to the set $\Sigma_{n^2}$.

Below we will give an algorithm (Algorithm \ref{Sigman^2}), which checks that the given binary matrix is $\Sigma_{n^2}$ and works in time $O(n^4 )$, i.e. $\theta_5  =O(n^4 )$ and therefore $z=4$.

Let $A_{st} =\left[\left(a_{st}\right)_{kl} \right]_{n\times n}$ be an arbitrary $n\times n$ matrix, $0\le k,l,s,t<n$. Let $A = \left[ \alpha_{ij} \right]_{n^2 \times n^2}$, $0\le i,j<n^2$ be an $n^2 \times n^2$ matrix such that for every $i,j,k,l, s,t$, $0\le i,j<n^2$ and $0\le k,l,s,t<n$, the equality  $\alpha_{ij} = \left(a_{st}\right)_{kl} $ is satisfied. Then, taking into account that the numbering of indexes starts from 0, it is easily seen that
\begin{equation}\label{ijks}
i=sn +k
\end{equation}
and
\begin{equation}\label{ijlt}
j=tn+l .
\end{equation}

\begin{algorithm}\label{Sigman^2}
Checking that a binary $n^2 \times n^2$ matrix $B=[\beta_{ij} ]\in \Sigma_{n^2}$. The algorithm gets the elements of the matrix $B=[\beta_{ij} ]$ from the array \verb"a[]" with size $n^4$, such that $\beta_{ij} = $\verb"a[i*n2+j]", where $\verb"n2"=n^2$, $0\le i<n^2 -1$, $0\le j<n^2 -1$.

For ease, we will not check for correctness of the input data, i.e. whether the input matrix is binary, in other words whether all its elements are 0 or 1. We leave this check to the reader for an exercise.

\begin{verbatim}
bool alg6(int a[], int n)
{
    int n2 = n*n;
    int r;
    int i,j,k,l,s,t;
    for (i=0; i<n2; i++)
    {
        r=0;
        for (j=0; j<n2; j++)
        {
            r = r+a[i*n2+j];
            if(r>1) return false; /* \end{verbatim}
\rm  There are more than one 1 in the i-th row.
\begin{verbatim} */
        }
        if(r==0) return false; /* \end{verbatim}
\rm  i-th row is completely zero.
\begin{verbatim} */
    }
    for (j=0; j<n2; j++)
    {
        r=0;
        for (i=0; i<n2; i++)
        {
            r = r+a[i*n2+j];
            if(r>1) return false; /* \end{verbatim}
\rm  There are more than one 1 in the i-th column.
\begin{verbatim} */
        }
        if(r==0) return false; /* \end{verbatim}
\rm  i-th column is completely zero.
\begin{verbatim} */
    }
    for (s=0; s<n; s++)
    {
        for (t=0; t<n; t++)
        {
            r=0;
            for (k=0; k<n; k++)
            {
                for (l=0; l<n; l++)
                {
                    i = s*n+k; /* \end{verbatim}
\rm  According to equation (\ref{ijks})
\begin{verbatim} */
                    j = t*n+l; /* \end{verbatim}
\rm  According to equation (\ref{ijlt})
\begin{verbatim} */
                    r= r+a[i*n2+j];
                }
            }
            if(r!=1) return false;
        }
    }	
    return true;
}

\end{verbatim}

\end{algorithm}

When we compare (\ref{ppp5}) with (\ref{p3(n)}) and (\ref{p4}), as well as (\ref{ttt5}) with (\ref{tau3}) and (\ref{tau4}) we may assume that algorithms which use random $\Pi_n$ matrices are expected to be more efficient (regard to probability and time for generating and checking)
than algorithms using random $\Sigma_{n^2}$ matrices to solve similar problems. This gives grounds for further examination of the $\Pi_n$ matrices' properties.

\section{Random Sudoku matrices}

We will give a little bit more complex definition of the term ''disjoint'' regarding $\Pi_n$ matrices. Let $C=\left[ c_{ij} \right]_{2n\times n}$ and $D=\left[ d_{ij} \right]_{2n\times n}$, be two $ \Pi_n$ matrices. We regard $C$ and $D$ as \emph{disjoint} matrices, if there are no natural numbers $s,t\in \{ 1,2,\ldots n\}$ such that the ordered pair $\langle c_{st} ,c_{n+t\; s} \rangle$ is equal to the ordered pair $\langle d_{st} ,d_{n+t\; s} \rangle$.

The relationship between $\Pi_n$ matrices and Sudoku matrices is illustrated by the following theorem, considering Proposition \ref{disj}.

\begin{theorem}\label{th1}
There is a bijective map from $\Pi_n$ to $\Sigma_{n^2}$ and the pair of disjoint matrices of $\Pi_n$ corresponds to the pair of disjoint matrices of $\Sigma_{n^2}$
\end{theorem}

Proof. Let $P=[\pi_{ij} ]_{2n\times n} \in \Pi_n$. We obtain an unique matrix of $\Sigma_{n^2}$ from $P$ by means of the following algorithm:

\begin{algorithm}\label{alg5}

From a given matrix  $M=\left[\pi_{ij} \right]_{2n\times n} \in \Pi_n$ we obtain a unique  matrix  $A= \left[ \alpha_{ij} \right]_{n^2 \times n^2}\in\Sigma_{n^2}$.  The algorithm gets the elements of the matrix $M=\left[\pi_{ij} \right]_{2n\times n}$ from the array \verb"p[]" with size $2n^2$, such that \verb"p[i*n+j]" $=\pi_{ij}$, $0\le i<2n$, $0\le j<n$ and obtained using Algorithm \ref{alg4}. The algorithm will write the elements of the matrix $A= \left[ \alpha_{ij} \right]_{n^2 \times n^2}$ in the array \verb"a[]" with size  $n^4$, such that $\alpha_{ij} = $\verb"a[i*n*n+j]", $0\le i,j<n^2$.

\begin{verbatim}
void alg7(int p[], int a[], int n)
{
    int i,j,s,t;
    int n2 = n*n;
    int n4 = n2*n2;
    for (i=0; i<n4; i++) a[i] = 0;
	
    int b[2][n][n];

    for (i=0; i<n; i++)
        for (j=0; j<n; j++) b[0][i][j] = p[i*n+j];
		
    for (i=0; i<n; i++)
        for (j=0; j<n; j++) b[1][j][i] = p[n2+i*n+j];
		
    for (s=0; s<n; s++)
        for (t=0; t<n; t++)
        {
            i = s*n+b[0][s][t]-1; // According to equation (20)
            j = t*n+b[1][s][t]-1; // According to equation (21)
            a[i*n2+j] = 1;
        }
		
}
\end{verbatim}

\end{algorithm}

For convenience, at the beginning of Algorithm \ref{alg5}, we constructed an auxiliary $2\times n\times n$ three-dimensional array \verb"b" such that $$\verb"b[0][i][j]"=\pi_{ij}$$
and
$$\verb"b[1][j][i]"=\pi_{n+i\, j}$$ is satisfied for each $i,j\in \{0,1,\ldots ,n-1\}$.

For example, from the matrix

\begin{equation}\label{primerb3}
P=\left[
    \begin{array}{ccc}
      1 & 3 & 2 \\
      2 & 3 & 1 \\
      1 & 2 & 3 \\
      3 & 1 & 2 \\
      1 & 2 & 3 \\
      1 & 2 & 3 \\
    \end{array}
  \right] \in \Pi_3 ,
\end{equation}
we get the array \verb"b". For more clearly, to the left of the comma we will write the elements of \verb"b[0]", and to the right of the comma -- the elements of \verb"b[1]":
\begin{equation}\label{masivb}
\verb"b" =
\left[
    \begin{array}{ccc}
      1,3 & 3,1 & 2,1 \\
      2,1 & 3,2 & 1,2 \\
      1,2 & 2,3 & 3,3 \\
    \end{array}
\right]
\end{equation}

From the array \verb"b", Algorithm \ref{alg5} obtains a binary matrix $A= \left[ \alpha_{ij} \right]_{n^2 \times n^2}$, $0\le i,j< n^2$ of type (\ref{matrA}) such that if $A_{s t} = \left[ \gamma_{ij}\right]_{n\times n}$, $0\le s,t< n$ is a block in $A$, then
\begin{equation}\label{aij=1}
\gamma_{ij} =
\left\{
\begin{array}{l}
  \verb"i = s*n+b[0][s][t]-1"  \\
  \verb"j = t*n+b[1][s][t]-1"
\end{array}
\right.
\end{equation}

From (\ref{aij=1})  follows that there is a single 1 in each block $A_{s t} \in A$.

Let $s\in \{ 0,1,\ldots ,n-1\}$. Since $s$-th row of the matrix $P$ is a permutation, then there is only one 1 in every row of $n\times n^2$ matrix
$$
R_s =
\left[
\begin{array}{cccc}
A_{s\, 0} & A_{s\, 1} & \cdots & A_{s\, n-1}
\end{array}
\right] .
$$

Similarly,  since ordered $n$-tuple $\langle p_{n+t\; 1} ,p_{n+t\; 2} ,\ldots ,p_{n+t\; n} \rangle$ which is $(n+t)$-th row of $P$ is a permutation for every $t\in \mathbb{Z}_n$, then there is only one 1 in every column of $n^2 \times n$ matrix
$$
C_t =
\left[
\begin{array}{c}
A_{1t}  \\
A_{2t} \\
\vdots \\
A_{nt}
\end{array}
\right] .
$$

Therefore, the matrix $A$ which is obtained with the help of Algorithm \ref{alg5} is a $\Sigma_{n^2}$ matrix.

For example, from the array B, which corresponds to the array (\ref{masivb}) and using Algorithm \ref{alg5}, we obtain the $\Sigma_9$-matrix
$$
A = \left[
\begin{array}{ccc|ccc|ccc}
      0 & 0 & 1 & 0 & 0 & 0 & 0 & 0 & 0 \\
      0 & 0 & 0 & 0 & 0 & 0 & 1 & 0 & 0 \\
      0 & 0 & 0 & 1 & 0 & 0 & 0 & 0 & 0 \\
      \hline
      0 & 0 & 0 & 0 & 0 & 0 & 0 & 1 & 0 \\
      1 & 0 & 0 & 0 & 0 & 0 & 0 & 0 & 0 \\
      0 & 0 & 0 & 0 & 1 & 0 & 0 & 0 & 0 \\
      \hline
      0 & 1 & 0 & 0 & 0 & 0 & 0 & 0 & 0 \\
      0 & 0 & 0 & 0 & 0 & 1 & 0 & 0 & 0 \\
      0 & 0 & 0 & 0 & 0 & 0 & 0 & 0 & 1 \\
    \end{array}
    \right]
    .
$$

Since a unique matrix  of $\Sigma_{n^2}$ is obtained for every $P\in \Pi_n$ by means of Algorithm \ref{alg5}, then this algorithm provides a description of the map $\varphi \; :\; \Pi_n \to \Sigma_{n^2}$. It is easy to see that if there are given different elements of $\Pi_n$, we can use Algorithm \ref{alg5} to obtain different elements of $\Sigma_{n^2}$. Hence, $\varphi$ is an injection. But according to formulas (\ref{Pi_n}) and (\ref{sssssss}) $\left| \Sigma_{n^2} \right| =\left| \Pi_n \right|$, whereby it follows that $\varphi$ is a bijection.

Analyzing Algorithm \ref{alg5}, we arrived at the conclusion that $P$ and $Q$ are disjoint matrices of $\Pi_n$ if and only if $\varphi (P)$ and $\varphi (Q)$ are disjoint matrices of $\Sigma_{n^2}$ according to the above mentioned definitions. The theorem is proved.

\hfill $\Box$

Let  $\mathcal{M}_{n^2} $ be the set of all square $n^2 \times n^2$ matrices with elements of the set $\mathbb{Z}_{n^2} =\{ 1,2,\ldots ,n^2 \}$, and let $\sigma_{n}$ be the number of all $n^2 \times n^2 $ Sudoku matrices. Obviously, $\left| \mathcal{M}_{n^2} \right| =(n^2 )^{n^2} =n^{2n^2}$. Then if we use Algorithm \ref{alg1} to obtain random Sudoku matrix. According to equation (\ref{p(n)}) there is the following probability evaluation:
\begin{equation}\label{p6}
p_6 (n) =\frac{\sigma_n}{n^{2n^2}}
\end{equation}

When $n=2$  $\sigma_2 = 288$  \citep{YORDZHEV20133072}.

When $n=3$, there are exactly
$$\sigma_3 = 6\; 670\; 903\; 752\; 021\; 072\; 936\; 960 \approx 6.671\times 10^{21}$$ in number Sudoku matrices \citep{Felgenhauer,AzBouki}.
As far as the author of this study knows, there is not a universal formula for the number $\sigma_n$ of  Sudoku matrices with every natural number $n$. We consider it as an open problem in mathematics.

If we employ random methods to create the matrix $P\in \mathcal{M}_{n^2}$ with elements of $\mathbb{Z}_{n^2}$, then according to Algorithm \ref{alg1} we need to verify if every row, every column and every block of $P$ is a permutation of elements of $\mathbb{Z}_{n^2}$. According to evaluation of Algorithm  \ref{alg2} , every verification can be done in time $O(n)$ (see equation (\ref{tau1})). Hence, when we employ Algorithm \ref{alg1} to obtain a random Sudoku matrix we will obtain the following time for generating and checking:
\begin{equation}\label{tau6}
\tau_6 (n) \in T(\mathcal{A}_n) \left( n^2 \right)^2 +2nO(n) +n^2 O(n) = t_0 \left( n^2 \right)^2 +2nO(n) +n^2 O(n) =O(n^4 ) .
\end{equation}

Here we will present a more efficient algorithm for obtaining random Sudoku matrix, based on the propositions and algorithms which are examined in the previous sections of this paper. The main point is to obtain $n^2$ in number random $\Pi_n$ matrices (Algorithm \ref{alg4}). For every $\Pi_n$ matrix which is obtained, it has to be checked if it is disjoint with each of the above obtained matrices.The criteria, described in Theorem \ref{th1} and Proposition \ref{disj} are used in the verification. If the obtained matrix is not disjoint with at least one of the above mentioned matrices, it has to be replaced with another randomly generated $\Pi_n$ matrix.

\begin{algorithm}\label{Sudokuend}
Obtaining random Sudoku matrix
\begin{verbatim}
#include <iostream>
#include <cstdlib>
#include <ctime>
#define N 3
#define N2 9   // N2 = N*N
#define N4 81  // N4 = N2*N2
#define d 18   // d = 2*N2
using namespace std;
int main()
{
    srand(time(0));
    int S[N4];
    int P[d];
    int A[N4];
    int z;
    for (int i=0; i<N4; i++) S[i] = 0;
    int K=1;
    while (K <= N2)
    {
        alg5(P,N);
        alg7(P,A,N);
        z=0;
        for (int i=0; i<N4; i++)
        {
            z += A[i]*S[i];
            if (z != 0) break;
        }
        if (z==0)
        {
            for (int i=0; i<N4; i++) S[i] += K*A[i];
            K++;
        }
    }
    for (int i=0; i<N4; i++)
    {
        cout<<S[i]<<" ";
        if ((i+1)%N2 == 0) cout<<endl;
    }
    return 0;
}
\end{verbatim}

\end{algorithm}

With the help of Algorithm \ref{Sudokuend}, we received a lot of random $9\times 9$ random Sudoku matrices, for example the next one:

$$
\left[
  \begin{array}{ccc|ccc|ccc}
    6 & 4 & 2 & 3 & 1 & 7 & 8 & 9 & 5 \\
    5 & 3 & 1 & 8 & 2 & 9 & 4 & 7 & 6 \\
    7 & 8 & 9 & 4 & 5 & 6 & 2 & 3 & 1 \\
    \hline
    9 & 6 & 7 & 2 & 4 & 5 & 1 & 8 & 3 \\
    3 & 2 & 4 & 6 & 8 & 1 & 9 & 5 & 7 \\
    1 & 5 & 8 & 9 & 7 & 3 & 6 & 2 & 4 \\
    \hline
    8 & 9 & 5 & 1 & 3 & 4 & 7 & 6 & 2 \\
    2 & 1 & 3 & 7 & 6 & 8 & 5 & 4 & 9 \\
    4 & 7 & 6 & 5 & 9 & 2 & 3 & 1 & 8 \\
  \end{array}
\right]
$$


\end{document}